\documentstyle[preprint, eqsecnum,aps]{revtex}

\tighten
\begin{document}

\draft
\preprint{HD-THEP-93-23 rev}
\title{Sphaleron effects near the critical temperature}
\author{Meik Hellmund}
\address{Institut f\"ur Theoretische Physik der Universit\"at
Leipzig}
\author{Jochen Kripfganz\thanks{Supported by Deutsche
 Forschungsgemeinschaft}
 and Michael G. Schmidt }
\address{Institut f\"ur Theoretische Physik der Universit\"at
Heidelberg}
\maketitle

\overfullrule0pt
\begin{abstract}
We discuss  one-loop radiative corrections to the
sphaleron-induced baryon number-violating transition rate
 near the electroweak
phase transition in the standard model. We emphasize that
in the case of a
first-order transition a rearrangement of the
loop expansion is required close to the transition
temperature.  The corresponding expansion parameter,
the effective 3-dimensional gauge coupling approaches a
 finite $\lambda$
dependent value at the critical temperature.
 The $\lambda$
 (Higgs mass) dependence of the 1-loop radiative corrections
is
discussed in the framework of the heat kernel method.
Radiative corrections are small compared to the leading
sphaleron contribution as long
as the Higgs mass is small compared to the W mass.
To 1-loop accuracy, there is no Higgs mass   range
compatible with experimental limits where  washing-out
of a B+L asymmetry could be avoided for the
minimal standard model with one Higgs doublet.

\end{abstract}
\pacs{ }

\narrowtext

\section{Introduction}

Baryon number is violated by nonperturbative effects
 in the electroweak
interaction \cite{1}. While it is under controversial
debate whether this could
show up  \cite{2,3} in high energy laboratory experiments,
there is a
common agreement that this should give important effects
at the
high temperatures of
the early universe \cite{4}. Indeed the discussion of
sphaleron
\cite{5} induced baryon number violation in a heat bath
triggered the work
on sizeable baryon number violation by the standard interactions.
Actually in the hot electroweak phase without Higgs
vacuum expectation value there is
no sphaleron configuration, but still washing out of
baryon number
is expected if $B-L=0$ (which is rather natural in inflationary
models without strong reheating). Thus we have the
difficult  problem
of explaining baryon number generation \cite{6} in a first-order
electroweak phase transition \cite{4,7}. Even if such a
mechanism can be convincingly described, one still has
to make sure
that the baryon asymmetry generated is not washed out
 by sphaleron effects
directly after the phase transition to the Higgs
phase. At these temperatures $T$ the effective potential
of the theory
should already produce a sizeable Higgs
vacuum expectation value $v(T)$ since a strong
first order transition is required for baryon number production.
This brings us to the subject of this paper, a discussion of the
wash-out by sphalerons with a transition rate \cite{4,8}

\begin{equation}\label{1.1}
\Gamma_{B\llap/}\sim\exp\left(-\frac{v(T)}{gT}S_3^{\text{Sphaleron}}\right)
\end{equation}

where $g$ is the gauge coupling and $S_3$ the 3-dimensional
sphaleron action. In the usual perturbative treatment of the
effective potential for the Higgs field $\phi$
\cite{9,10,11,32,16,33} $v(T)\not=0$ is
directly related to a radiatively generated
$(\phi^+\phi)^{3/2}T$-term.
It also depends
on the Higgs coupling $\lambda$ and hence is sensitive to
the Higgs mass. Requiring the baryon asymmetry not being washed
out by the sphaleron effect (eq.~(\ref{1.1})), an upper bound to the
Higgs mass can be derived \cite{12}. The rough bound $m_H<45$
 GeV
is already significantly violated by experiments.

  Radiative corrections
\cite{8,14,15}, the factor in front of eq.~(\ref{1.1}),
 have to be discussed. If perturbation theory in the effective
3-dimensional gauge
coupling is reliable the conclusion remains unaltered that
the
standard electroweak theory leads to a washing out of a  previously
generated baryon
asymmetry (if $B-L=0$) after the phase
transition. Because of the magnitude of the effective coupling, this
 perturbative expansion is problematic
 for $m_H$ near $M_W$ close to the phase transition, whereas for
smaller values of $m_H$ it is more trustworthy. In any case,
nonperturbative effects  could be responsible for a stronger first
order transition, as suggested by numerical lattice studies
\cite{29,31}.
  If the problem of washing out the baryon asymmetry could be avoided,
CP violation might be strong enough to explain the baryon number of
the universe even in the standard model
with a single doublet
 \cite{35}. Otherwise,   variants of the
standard electroweak theory would have to be considered  \cite{7,13}.
Also in these
models, however, a thorough discussion of the baryon number
violating rate
including radiative effects is necessary if one wants to give quantitative
bounds for the Higgs mass.

One might doubt whether the perturbative treatment of the effective
action in the temperature range close to the phase transition and
with field configurations restricted by the sphaleron
ansatz is reliable. There are several dangerous aspects we should
shortly mention  at the beginning and comment on in more detail later
 on.

 The region  around the Higgs field $\phi=0$ has strong infrared
effects
like those in QCD and one has to inspect how sensitive to this
region the
sphaleron calculation is. As argued in ref. \cite{16}, infrared
divergences
due to Goldstone bosons  also
appear for $\phi\sim v(T)$
 in the selfconsistency (gap type) equations if the Landau gauge is used.
We do not expect these singularities to be physical.

 The sphaleron is a 3-dimensional object, and it is consequent to
discuss it with an effective action where
the nonzero Matsubara frequencies are already integrated out, i.\ e. in
 a 3-dimensional
 theory for the static modes. In low order perturbation theory,
combined with a high temperature expansion
this can be
carried out. The non-static modes are expected to behave
 perturbatively, except for
plasma mass effects,
whereas the static (zero) modes may show non-perturbative behaviour
(dynamical
mass generation, condensate formation, confinement). It is convenient to
use the high temperature expansion in integrating
the non-static modes. However, the phase transition may be at
temperatures where
this is not valid anymore.

It is only partly a matter of convenience which radiative corrections
should
already be included in the effective action which gives the sphaleron
configuration.
In principle one could use ``exact'' renormalization group
techniques \cite{19,20} to arrive at the appropriate action at some
scale, but even the discussion of one-loop corrections to the
fundamental Lagrangian with the well-known sphaleron gives interesting
insights.

 The 3-dimensional effective (dimensionless) gauge coupling
$g^2_3(T)=\frac{g\cdot T}{v(T)}$
is not very small at the temperatures of the supposed
first order phase
transition (but does not diverge as for second-order transitions).
This supports the high temperature expansion, but causes problems
with the perturbative expansion.
Though it would be
certainly more satisfying to see all these aspects controlled in
a genuinely nonperturbative treatment, a perturbative approach might still
give a good orientation.

When this paper was first written \cite{42}, an exact calculation of the
1-loop sphaleron
determinant by  Carson et al. \cite{15} had been available for relatively
large values of
the Higgs mass.  The result apparently differed considerably from the
 answer one would get in determining the sphaleron using the 1-loop
effective potential.  Later,  Baacke and Junker
\cite{43}
presented another complete calculation of the determinant, with rather
different results.
In the mean-time, some errors have been corrected, and the use of
different schemes has been clarified by these authors \cite{44}.  Both
'exact' answers are now essentially consistent,
and are furthermore surprisingly well represented up to intermediate
Higgs masses
by  the  $\phi^3$ term of the 1-loop effective potential.
The results of refs. \cite{15,44}
cannot directly be used near the electroweak phase transition
because the
original expansion parameter becomes infinite. The loop expansion has
 to be rearranged.
The $\lambda$ dependence of the 1-loop
corrections will be estimated using the heat-kernel expansion.  These
 will be the main subjects of this paper.

Chapter 2 contains a discussion of the effective
three-dimensional action including the $\phi^3$-term at temperatures close
to the phase transition. In chapter 3
we study the radiative
corrections to the  baryon number violating transition rate induced by the
sphaleron. The $\lambda$ (Higgs mass) dependence of
the different parts of the rate are discussed in detail.  Using the
't Hooft-Feynman gauge plays
an important role.
Chapter 4 contains numerical evaluations and
a discussion of our results.

\section{The one-loop effective action near the phase transition}

The effective action for the sphaleron saddle point \cite{5}
can differ from the fundamental action. The  inclusion of
1-loop effects (eventually enlarged by the selfconsistent
 IR plasma mass corrections)
has important effects in temperature
quantum field theory, in particular the $T^2\phi^2$
plasma mass term. There is also a $T$-dependent coupling
$\lambda_T$ and there
will appear new terms in the action like $\phi^6/T^2$ etc. if one completely
integrates our the nonzero Matsubara frequencies. The latter terms are
neglected in the following in the spirit of a high temperature expansion.

After a rescaling

\begin{equation}\label{2.1}
r\to\frac{\xi}{gv_0(T)},\quad A_\mu\to v_0(T) A_\mu,\quad
\phi\to v_0(T)\phi
\end{equation}

the electroweak standard action (without Weinberg mixing)
including plasma mass
and $\lambda_T$, and
reduced to three dimensions
can be written as \cite{14,15}

\begin{eqnarray}\label{2.2}
\frac{1}{g^{(0)2}_3(T)}S^{(0)}_3&=&\frac{1}{g^{(0)2}_3(T)}\int d^3\xi\Bigl[
\frac{1}{4}F^a_{ik}F^a_{ik}+(D_i\phi)^+(D_i\phi)\nonumber\\
& &+\frac{\lambda_T}{g^2}\left(\phi^+\phi-\frac{1}{2}\right)^2\nonumber\\& &
+\frac{1}{2}A_0\left(-D_iD_i+
\frac{1}{2}\phi^+\phi \right)A_0\Bigr]
\end{eqnarray}

with the effective 3-dimensional gauge coupling

\begin{equation}\label{2.3}
g^{(0)2}_3(T)=\frac{gT}{v_0(T)}
\end{equation}

and with T-dependent minimum of the potential
(in the notation of ref. \cite{9})

\begin{equation}\label{2.4}
v_0^2(T)=\frac{2}{\lambda_T}(T^2_0-T^2)D
\end{equation}

where

\begin{eqnarray}\label{2.5}
D&=&\frac{1}{8v^2_0}(3m^2_W+2m_t^2)\qquad (v_0=246 GeV),\nonumber\\
T^2_0&=&\frac{m_H^2-8v_0^2B}{4D}\nonumber\\
m^2_W&=&\frac{1}{4}g^2v_0^2,\quad m_H^2=2\lambda v^2_0\nonumber\\
\lambda_T&=&\lambda-\frac{3}{16\pi^2v^4_0}\left(3M_W^4\log\frac{M_W^2}
{a_BT^2}-4m^4_t\log\frac{m^2_t}{a_FT^2}\right)\nonumber\\
B&=&\frac{3}{64\pi^2v^4_0}(3m^4_W-4m_t^4)\nonumber\\
\log a_B &=&2\log{4\pi} -2\gamma\nonumber\\
\log a_F &=&2\log{\pi} -2\gamma
\end{eqnarray}

This would predict a second-order phase transition at $T\to T_0$, where $v_0
(T)\to 0$ and hence $g_3^{(0)2}(T)\to\infty$.
The saddle point approximation would break down close to $T_0$, as well as the
perturbative expansion in $g^{(0)2}_3$.
Discussing the baryon asymmetry  a first order phase transition is required
and indeed perturbation theory provides us with a term
$\sim T(g^2\phi^+\phi)^{\frac{3}{2}}$
in the effective potential suggesting a first-order phase
transition.
This might be as misleading as in the pure $\phi^4$ theory where the coupling
$\lambda$ appearing in this term vanishes \cite{23} at the phase transition.
However, the latter does not happen with the 4-dimensional gauge coupling $g$.
The term above is a
zero-mode 3-dimensional IR effect

\begin{equation}\label{2.6}
V(m^2)\sim(m^2)^{\frac{3}{2}}T
\end{equation}

with $m^2=\frac{g^2\phi^+\phi}{2}$ for gauge bosons.

There are of course further contributions to the effective
action~(\ref{2.2}), among them a Debye mass term

\[
\frac{1}{2}(\frac{1}{6}(5+N_F)g^2T^2)A^a_0A^a_0
\]

Because of this
longitudinal gauge boson plasma mass$^2\sim T^2$ their
contribution is suppressed
at high temperatures.
There is also a mass counter term depending linearly on the cut-off. It arises
because the 4-dimensional one-loop integrals leave out the static modes
and cancel against divergences of the 3-dimensional theory.

Thus we include a term $-E(2\phi^+\phi)^{\frac{3}{2}}$ in the  potential
with

\[
E=\frac{2}{3}\frac{3}{32\pi}g^3
\]

where the factor $\frac{2}{3}$ is due to the suppression
 of the
longitudinal gauge boson in eq.~(\ref{2.2})  \cite{9} .
Following the discussion \cite{9,10} this leads to a
first-order transition to a new minimum

\begin{equation}\label{2.7}
v(T)=\frac{3}{2}\frac{ET}{\lambda_T}+\sqrt{\left(\frac{3}{2}\frac
{ET}{\lambda_T}\right)^2+v_0^2(T)}
\end{equation}

where $v_0(T)$ is given in eq.~(\ref{2.4}).
The critical temperature where both minima
of the potential are equal is at

\begin{equation}\label{2.8}
T^2_c=\frac{T_0^2}{1-\frac{E^2}{\lambda_{T_c}D}}
\end{equation}

where $v(T_c)=\frac{2ET_c}{\lambda_{T_c}}$, and
classical instability at $\phi=0$ sets
in at $T=T_0$ where $v(T_0)=\frac{3ET_0}{\lambda_{T_0}}$.

If we rescale with $v(T)$ instead of $v_0(T)$, we obtain an
effective action like~(\ref{2.2}) with

\begin{equation}\label{2.9}
g_3^2(T)=\frac{gT}{v(T)}
\end{equation}

but with a modified effective potential

\begin{eqnarray}\label{2.10}
V_{\text{eff}}&=&\frac{1}{g^2_3(T)}
\frac{\lambda_T}{g^2}\int d^3\xi\left[\left(\phi^+\phi-\frac{1}{2}
\right)^2+
\epsilon\left(\frac{3}{4}\left(\phi^+\phi-\frac{1}{2}\right)
\right.\right.
\nonumber\\
& &-\left.\left.\frac{1}{\sqrt{2}}
\left((\phi^+\phi)^\frac{3}{2}-(\frac{1}{2})^{3/2}\right)\right)\right].
\end{eqnarray}

The second part is related to the first one by a new parameter $\epsilon$, on
which
the quasiclassical solution will depend.

\begin{equation}\label{2.11}
\epsilon(T)=\frac{4}{3}\left(1-\frac{v^2_0(T)}{v^2(T)}\right)=
\frac{8}{3+\left(1+8\frac{T^2_0(T^2_c-T^2)}{T^2(T^2_c-T^2_0)}
\right)^\frac{1}{2}}
\end{equation}

which is $\epsilon=2$ for $T=T_c,\epsilon=\frac{4}{3}$ for $T=T_0$, and
(formally) $\epsilon=0$ for $T=0$. Thus in the region of the phase transition
$\epsilon\sim 0(1)$ the second term has a prefactor of
similar size as the first
one. There are two expansion parameters
$\frac{g^2_3(T)}{4\pi}$ and $\frac{\lambda_T}
{g^2}$,  varying independently through  T and $\lambda$. According to

\begin{equation}\label{2.12}
\frac{g^2_3(T)}{4\pi}=\frac{\lambda_T}{g^2}\frac{g^3}{16\pi E}\epsilon=
\frac{\lambda_T}{g^2}
\epsilon
\end{equation}

they are of about equal size
close to the phase transition.  The effective gauge coupling $g^2_3(T)$
is kept small with small $\frac{\lambda_T}{g^2}$, which is the case for small
Higgs
mass.   However, due to a heavy top quark , $\frac{\lambda_T}{g^2}$ cannot be
arbitrarily  small. For $m_{top}=150 GeV$ and $m_{top}=200 GeV$ it is bounded
from below by 0.032 and 0.08, resp., following  eq.~(\ref{2.5}).  The relation
between
$\frac{\lambda_T}{g^2}$ and $\frac{m_H^2}{M_W^2}$
at $T=T_0$
is shown in Fig.\ \ref{fig1}.

In the case $m^2_H\sim M_W^2$  $g^2_3(T_c)$ is of order one, and it is not
clear a priori whether this is small enough for the quasiclassical expansion
(in the broken symmetry phase) to be well behaved.

The high temperature expansion breaks down
if $g^2_3(T)$ is too small:

\begin{equation}\label{2.13}
\frac{M_W(T)}{T}=\frac{gv(T)}{2T}=\frac{g^2}{2g^2_3(T)}=\frac{2g^2}
{(\lambda_T/g^2)}
\frac{1}{16\pi\epsilon}
\end{equation}

This can only be avoided in the limit $g\to0,\frac{\lambda_T}{g^2}$ fixed.
Assuming a limit of validity of the high temperature expansion of
$\frac{M_W(T)}{T}<2,$
eq.~(\ref{2.13}) requires $g_3^2(T)>\frac{g^2}{4}$.
For a large top mass the requirement $\frac{m_t(T)}{T}<2$ is more restrictive
but in this case an exact treatment
of the temperature dependence is no problem.

The 3-dimensional reduction based on the dominance of zero modes, crucial
for the $\phi^3T$-term requires $2\pi T\gg M_W(T)$ or $\pi T>m_t(T)$,
respectively,
which is less restrictive than the previous inequality.

\section{One-loop fluctuations around the sphaleron}

The transition rate \cite{22,24} induced by the sphaleron (responsible
for a possible wash-out of a baryon asymmetry) is \cite{8,14,25,37}

\begin{eqnarray}\label{3.1}
\Gamma/V&=&\frac{\omega_-}{2\pi}
N_{\text{trans}}(NV)_{\text{rot}}\left(\frac{g^2T}{(4\pi)^2}
\right)^3\left(
\frac{g^2_3}{4\pi}\right)^{-6}
\nonumber\\
& &\times
\exp\left(-\frac{1}{g^2_3(T)}S^{(0)}_3-
S^{(1,\alpha)}_3-S_3^{(1,\beta)}\right)
\end{eqnarray}

 Here $\omega^2_-$ is the  eigenvalue of the unstable mode
and as usual the integration
over collective coordinates corresponding to translational and rotational zero
modes is taken out explicitly. $S^{(0)}_3$ is
the classical
sphaleron action, and $S_3^{(1,\alpha)}$ that part of the one-loop
action obtained from the
$\phi^3$ effective potential and the corresponding shift of $g^2_3(T)$

\begin{equation}\label{3.2}
S_3^{(1,\alpha)}=\tilde E\int d^3\xi
\left(\frac{3}{4}(\phi^+\phi-\frac{1}{2})-
\frac{1}{\sqrt{2}}((\phi^+\phi)^{\frac{3}
{2}}-(\frac{1}{2})^{3/2})\right)
\end{equation}

with $\tilde E=\frac{4E}{g^3}=\frac{1}{4\pi}$ according to eq.~(\ref{2.12}).
$S_3^{(1,\beta)}$ contains the remaining 1-loop contribution, i.~e.
the full 1-loop
action with the $\phi^3$ term subtracted

\begin{equation}\label{3.3}
S_3^{(1,\beta)}=-(\log\kappa-\frac{\tilde E}{\sqrt2}
\int d^3\xi((\phi^+\phi)^{\frac{3}{2}}-(
\frac{1}{2})^{3/2})
\end{equation}

$\kappa$ is given by

\begin{equation}\label{3.4}
\kappa=\text{Im}\left(\frac{\det'\left(
\frac{\delta^2S_{gf}}{\delta\phi^2}\right)_{\phi=\phi_{vac}}}
{\det'\left(\frac{\delta^2S_{gf}}{\delta\phi^2}
\right)_{\phi=\phi_{sp}}}\right)^{1/2}
\left(\frac{\Delta_{FP\phi=\phi_{sp}}}{\Delta_{FP\phi=\phi_{vac}}}\right)
\end{equation}

We follow the notation of ref. \cite{14,15}. The prime in eq.~(\ref{3.4})
indicates that zero modes have been
removed from the sphaleron determinant and a corresponding number of low-lying
eigenvalues are dropped in the free field determinant. $\kappa$ still contains
the unstable mode contribution, which also
has to be taken out of the determinant
explicitly. We discuss this fluctuation determinant in the framework of the
heat-kernel expansion.
In the Schwinger proper time formulation one has

\begin{equation}\label{3.5}
\log\frac{\det K}{\det K_0}=
\lim_{\epsilon\to 0}\left(-tr\int^\infty_\epsilon\frac{dt}{t}
(e^{-tK}-e^{-tK_0}\right)
\end{equation}

At large $t$,  the integrand is well approximated by the lowest eigenstates in
the spectrum. For $m^2_H\sim m^2_W$ there is only one scale in the
theory, both in the vacuum sector (``$K_0$'') and the sphaleron sector
(``$K$''), and the large $t$ behaviour can be well parametrized by a
form $Ae^{-tm^2_W}$ after subtracting  the zero modes and  the unstable
mode which have to be taken out explicitly. In numerical
evaluations \cite{14} the subtraction of the exponentially
increasing mode contribution is very problematic and deserves  further
studies.

Because
of the exponential damping, the main part of the integral comes from
%$t\,m^2_W< const.$
small $t\,m^2_W$
where an expansion of the exponentials in~(\ref{3.5}) in $t$
makes sense, though the radius of convergence is not known.
Such a t expansion has
been carried out  in  ref. \cite{14}.
The expression (3.5) can be transformed into

\begin{eqnarray}\label{3.6}
\log\frac{\det K}{\det K_0}&=&-\int^\infty_0\frac{dt}{t}
\int d^dx\int\frac{d^dp}{(2\pi)}
e^{-tp^2}\nonumber\\
& &\times \text{tr}(e^{-t\delta K}-e^{-t\delta  K_0})
\end{eqnarray}

with

\begin{eqnarray*}
\delta K(p,x)&=&-2ip\cdot D(A)-(D(A))^2+V\\
\delta K(p,x)&=&-2ip\partial-\partial^2+V_0
\end{eqnarray*}

In the present case,
an explicit expansion in $t$ can be obtained.  $V$ is a $13\times 13$
matrix
(or a $3\times 3$ matrix for the ghosts).
After $p$-integration, the authors of ref.
\cite{14} find $(d=3)$

\begin{equation}\label{3.7}
\log\frac{\det K}{\det K_0}=-\int^\infty_0
\frac{dt}{t}\frac{1}{(4\pi t)^{d/2}}\sum
^\infty_{p=1}\frac{(-t)^p}{p!}O_p
\end{equation}

with the first three operators $O_p, p=1,2,3$
 given explicitly in ref. \cite{14}. Since we use a different rescaling, our
fluctuation operator differs from that of ref. \cite{14} in replacing
$(\phi^+\phi-\frac{1}{2})$ by $(\phi^+\phi-\frac{1}{2}\frac{v_0^2}{v^2})$.
This leads to
corresponding modifications of prefactors of derivative operators as well.

The expansion~(\ref{3.7})
is appropriate for the small $t$ part of the integral. Eq. (3.7)
should therefore be applied to some  $t$-interval $(0,\bar t)$ only.
$\bar t$ should
be chosen to smoothly approach the large $t$ behaviour governed by the first
few eigenvalues of $K$ and $K_0$, respectively.

Convergent $t$ integrals are obtained if the exponential $e^{-m^2_0t}$, with
$m_0$ the mass gap of the free theory (i.\ e. $K_0$) is not expanded
but taken out. This leads to some operator rearrangements
$O_p\to\tilde O_p$.  This exponential fall-off does not properly
describe the behaviour of $e^{-Kt}$, however, which is governed by the
unstable mode,  various zero modes, and possible bound states at
positive energy (studied in ref. \cite{27}).  After proper subtractions
the first positive eigenvalue \cite{27} of $K$ should dominate.

It is tempting to try to sum up subsets of operators to all orders. The most
prominent case are operators with no derivatives.
If in eq.~(\ref{3.6}) all derivative operators are dropped, one obtains
\widetext
\begin{equation}\label{3.8}
\log\frac{\det K}{\det  K_0}=I(m^2)-I(m^2_0)
=-\int^\infty_0\frac{dt}{t}
\int d^3x\int \frac{d^3p}{(2\pi)^3}e^{-tp^2}(e^{-m^2t}
-e^{-m^2_0t})
\end{equation}

First performing the $p$ integral one finds

\begin{equation}\label{3.9}
I(m^2)-I(m^2_0)=-\int\frac{dt}{t}\int d^3x\frac{1}{(4\pi t)^{3/2}}e^{-m^2_0t}
(e^{(m^2_0-m^2)t}-1)
\end{equation}

Expanding the last factor in t and performing the $t$ integration leads to

\begin{equation}\label{3.10}
I(m^2)-I(m^2_0)=-\left(\frac{m^2_0}{4\pi}\right)^{3/2}\sum^\infty_{p=1}
\frac{\Gamma(p-3/2)}{\Gamma(p+1)}
\int d^3x\left(\frac{m^2_0-m^2}{m^2_0}\right)^p
\end{equation}

\narrowtext

This procedure automatically regularizes the UV divergence originally present
(for $p=1$).

Eq.~(\ref{3.6}) can of course be evaluated directly. First performing the $t$
integral and differentiating twice with respect to $m^2$ in order to
regularize the UV
diagram, we find

\begin{eqnarray}\label{3.11}
I''(m^2)&=&-\int d^3x\int\frac{d^3p}{(2\pi)^3}\frac{1}{(p^2+m^2)^2}
\nonumber\\
&=&-\frac{1}{8\pi}
\int d^3x\,\frac{1}{\sqrt{m^2}}
\end{eqnarray}

Integrating again produces the $\phi^3$ term

\begin{equation}\label{3.12}
I(m^2)-I(m^2_0)=-\frac{1}{2\pi}\frac{1}{3}
\int d^3x\,\left( (m^2(\phi))^{3/2}-(m^2_0)^{3/2}\right)
\end{equation}

Expanding around $m^2=m^2_0$ leads to eq. (3.10).
Uncertainties due to
finite renormalizations do not arise here because the
linear divergence is not present
in the full (4-dim) theory, and therefore no corresponding
counterterms are required.

The correct normalization
of the $\phi^3$ term is obtained by counting 9 degrees
of freedom for W's,
three for Goldstone bosons, one for the Higgs boson, and
subtracting six for the ghosts
with masses

\begin{eqnarray}\label{3.13}
m^2_W&=&\frac{1}{2}\phi^+\phi\nonumber\\
m^2_\chi&=&\frac{1}{2}\phi^+\phi+\frac{2\lambda}
{g^2}
\left(\phi^+\phi-\frac{1}{2}\frac{v_0^2(T)}{v^2(T)}\right)\nonumber\\
m^2_H&=&\frac{2\lambda}{g^2}\left(3\phi^+\phi-\frac{1}{2}
\frac{v_0^2(T)}{v^2(T)}\right)\nonumber\\
m^2_{gh}&=&\frac{1}{2}\phi^+\phi
\end{eqnarray}

respectively. In the 't Hooft-Feynman gauge the Goldstone
boson is massive  in the
vacuum of the broken symmetry phase, in contrast to
Landau gauge (and $v=v_0$).

The adjoint Higgs representation of the dimensionally
reduced theory
(corresponding
to the $A_0$ field) is not taken into account because
of plasma masses.
Therefore, we do not find the partial cancellation between
ghosts and
the adjoint Higgs field (i.\ e. $A_0$) assumed in ref. \cite{14}.
In this way we
get the properly reduced coefficient of the $\phi^3$ term.

If we keep the full $\lambda$ dependent masses eq.(3.13)
 we generate the complete
1-loop effective potential and not just the $\phi^3$ term

\begin{eqnarray}
V_{\text{eff}}&=&
\frac{1}{g_3^2}\int d^3\xi\left[\frac{\lambda}{g^2}\left(
(\phi^+\phi)^2-\frac{v_0^2}{v^2}\phi^+\phi\right)\right.\nonumber\\
& &-g_3^2\left\{\frac{1}{4\pi}\left(\frac{1}{2}\phi^+\phi\right)^{3/2}
\right.\nonumber\\
& &+\frac{1}{4\pi}\left(\frac{1}{2}\phi^+\phi+\frac{2\lambda}{g^2}
\left(\phi^+\phi-\frac{1}{2}\frac{v_0^2}{v^2}\right)\right)^{3/2}\nonumber\\
& &\left.\left.
+\frac{1}{12\pi}\left(\frac{2\lambda}{g^2}\left(3\phi^+\phi-\frac{1}{2}
\frac{v_0^2}{v^2}\right)\right)^{3/2}\right\}\right]
\end{eqnarray}

In our notation, the difference to the $\phi^3$ term is one
contribution to $S^{(1,\beta)}$

\begin{eqnarray}
S_V^{(1,\beta)}&=&
\int
d^3\xi %\left[
%\right.\nonumber\\ & &
\left\{\frac{1}{4\pi}\left(\frac{1}{2}\phi^+\phi\right)^{3/2}
%\right.
\right.\nonumber\\
& &-\frac{1}{4\pi}\left(\frac{1}{2}\phi^+\phi+\frac{2\lambda}{g^2}
\left(\phi^+\phi-\frac{1}{2}\frac{v_0^2}{v^2}\right)\right)^{3/2}\nonumber\\
& &\left.
%\left.
-\frac{1}{12\pi}\left(\frac{2\lambda}{g^2}\left(3\phi^+\phi-\frac{1}{2}
\frac{v_0^2}{v^2}\right)\right)^{3/2}\right\} %\right]
\end{eqnarray}

to be subtracted at $\phi^+\phi = \frac{1}{2}$.

This contribution is well behaved above $T_0$. Below, it becomes
 complex.  This
already tells us that one should be reluctant in summing partial sets
 of operators.

One can  attempt to sum further  particular classes of operators, with
a given number of derivatives.
This would generate the derivative expansion of the effective action.
The effective
potential has already been discussed. As one nontrivial example we
present the
contribution of the field strength operator $FF$ to $S^{(1,\beta)}$

\begin{equation}\label{3.14}
S_{FF}^{(1,\beta)}=-\int d^3\xi\left(\frac{173}{24}\right)\frac{1}{32\pi}
\left(\frac{\phi^+\phi}{2}\right)^{-1/2}F_{ij}^aF_{ij}^a
\end{equation}

obtained from summing up the $\lambda$-independent parts of
 $(\phi^+\phi)^nFF$, arising from the gauge, Goldstone, and ghost loop,
resp. A numerical discussion will be given in chapter 4.

It is instructive to keep the $t$ integral in analogy to eq.~(\ref{3.7}).
 In a sphaleron
background, the integrand
 falls off only power-like with t. These partially
summed operators therefore represent a $t$-dependence very different from the
true large $t$ behaviour. This causes some doubts on whether these
contributions separately give a good representation of the complete effective
action.

Considering high derivative operators one actually finds the  t integrand
increasing
power-like in t, i.\ e. the t-integrals diverge, and
 the derivative expansion actually breaks
down for the sphaleron. The reason is that  $m^2(\phi)$ approaches zero for
$\xi\to0$
in the sphaleron background. Therefore, the corresponding corner
of the integrand is not exponentially suppressed in t.
This could be interpreted as being  related to the presence of the negative
mode
which could be build up more and more by higher derivative terms.

Partial resummation of certain operators becomes meaningful, however,
in studying the leading contributions at small $\lambda$. These singularities
 arise from the large $\xi$ behaviour
of the $\xi$-integral, because of the slow fall-off of the Higgs component
of the sphaleron at small $\lambda$. At small $m_H$, the sphaleron has
a core with size $O(m^{-1}_W)$.
Outside the gauge field vanishes whereas the Higgs
field falls off only like $e^{-m_Hr}$. More precisely, the asymptotic behavior
is

\begin{equation}\label{3.15}
\sqrt{2}|\phi|\;\;{\mathrel{\mathop
\simeq_{r\to\infty}}}\;\; 1-\frac{const.}{r}e^{-m_Hr}
\end{equation}

For $m_H\to 0,$ $S_3^{(1,\alpha)}$ will be singular because of the
large $r$ behaviour. However,
the leading $\frac{1}{\lambda_T}$ singularity cancels between
the two contributions.
Therefore,  $S_3^{(1,\alpha)}$ has only a square root singularity
in $\frac{1}{\lambda_T}$ ,
in contrast to the $\frac{1}{\lambda_T}$ singularity of the $\phi^3$ term.

Operators containing the gauge field
strength $F_{\mu\nu}$ fall off sufficiently fast, and also terms
containing two or more covariant derivatives $D_\mu\phi$ lead to
convergent
integrals. The
powerlike singular contributions to the effective action are therefore just
given by the effective potential.

$\lambda$ singularities associated with ``bound states''
 in the fluctuation spectrum would
only be logarithmic. If $m_+$ denotes such a bound state mass which
 vanishes with
$\lambda$ the corresponding contribution from the
large t region is proportional to $\log\frac{m^2_+}{\bar t}$.
 The $\bar t$-dependence
will cancel against contributions from the small t region,
but a logarithmic term
$\log\frac{m^2_+}{m^2_W}$ could remain. Stronger $\lambda$
singularities can only
arise from the small t region and should show up in the heat kernel expansion.

\section{Discussion}

Necessary criteria for a well-behaved quasi-classical treatment of sphaleron
transitions
is that 1-loop terms are small compared to the classical sphaleron action.
In the original
formulation this was not the case because the coupling blows up at the critical
temperature. This is cured by redefining the coupling according to the one-loop
vacuum expectation value $v(T)$, as outlined in the  chapter before.
The question now is whether one-loop terms
will be small compared to $\frac{1}{g^2_3}S_0$ even at the critical
temperature.
% if  $m_H$ (i.e. $\lambda/g^2$) is not too large.

 Singularities in $\lambda$
arise because some of the operators appearing in the expansion
become singular for $\lambda$ going to zero.
Those operators are just the ones without
derivatives. They are either multiplied with powers of $\lambda$,
and  therefore
are actually non-singular, ore can be summed up to
produce the $\phi^3$ term, i.\ e.
cancel in $S_3^{(1,\beta)}$. To any finite order in the heat-kernel
expansion,
$S_3^{(1,\beta)}$ will be nonsingular in $\lambda$. A separation
of the 1-loop correction into $S_3^{(1,\alpha)}$ and
 $S_3^{(1,\beta)}$ is meaningful because of the different
singularity structure in $\lambda$.

In Table\ \ref{tab}, we present numerical results for all the operators
 occuring in the
heat kernel expansion up to third order, i.\ e. as far as the
expansion has been worked
out in ref. \cite{14}.  This numerical study confirms that only the
first few non-derivative operators show a strong  $\lambda$ dependence.

To this low order in the heat kernel expansion it is not possible to give a
reliable estimate of $S_3^{(1,\beta)}.$ In particular, introducing an
exponential cut-off following ref. \cite{38} does not lead to stable
results.  For reliable quantitative statements on the absolute
normalization of
 $S_3^{(1,\beta)}$ it will be necessary to work out more terms
of the heat kernel
expansion. This presumably requires the development of new
techniques \cite{41}.

Since we can isolate those operators in the heat kernel expansion
showing a strong $\lambda$ dependence we should understand
the $\lambda$ dependence of  the sphaleron rate. A complete one-loop
calculation at one particular $\lambda$ value (e.\ g.\ $m_H=M_W$)
could thus
be used to normalize the rate, and predict the $\lambda$ dependence
through our results.

We now present a numerical study of  1-loop corrections to the
sphaleron rate.
For a quick estimate, we can simply replace the Higgs potential
by the 1-loop effective
potential in studying  the
sphaleron solution. The resulting sphaleron
energy is shown in Fig.~\ref{fig2}, in terms of the profile function
$A(\lambda_T/g^2,\epsilon)$
defined by

\begin{equation}\label{4.1}
\frac{E_{sp}}{T}=\frac{2\pi}{g^2_3(T)}A(\lambda_T/g^2,\epsilon(T))
\end{equation}

with $\epsilon(T)$ given by eq. (2.11). Even at the critical temperature,
i.\ e. $\epsilon=2$, $A$ changes not more than about 10 \%. The
$T$-dependence of the sphaleron transition rate is therefore almost
exclusively given by the behaviour of $g^2_3(T)$. Similar plots have
been given independently in
ref. \cite{26}.

For a more systematic study, one should not work with such a
'deformed' sphaleron but present the one-loop correction in terms
of the original sphaleron solution.
Results are shown in Fig.~\ref{fig3}, as function of $\lambda_T/g^2$. The
solid line represents the leading order contribution at $T=T_c$. It
is only defined after reordering the loop expansion by absorbing
appropriate 1-loop contributions into $g_3^2$. If  this would not
be done, the leading term would be zero at $T=T_0$, and the
one-loop correction would essentially be given by the $\phi^3$ term
 (dashed-dotted line in Fig.~\ref{fig3}) in the $\lambda$ range considered.
It is amusing to note that the two procedures would predict a
quite similar sphaleron rate. However, in the latter case the
 prediction is meaningless because the expansion parameter
 becomes infinite.

In our scheme, the $\phi^3$ term does not give the leading
1-loop correction but is subtracted by a $\phi^2$ term with
the same singularity in $\lambda$. The resulting contribution
 $S_3^{(1,\alpha)}$
(shown as dashed line in Fig.~\ref{fig3}) is very small compared
to the leading term in the whole $\lambda$ range. This is
not the case for the remaining 1-loop corrections.
 $S_V^{(1,\beta)}$ is shown as dotted line in Fig.~\ref{fig3}.
 It is small at small $\lambda$ but increases and becomes
 of the order of the leading term for $\lambda$ larger than
 about 0.1.  $S_{FF}^{(1,\beta)}$ shows a similar behaviour,
presented in Fig.~\ref{fig4}. Again, the correction becomes of order
 one for $\lambda_T$ about 0.1. Its sign is different from
that of  $S_V^{(1,\beta)}$, however. There are two possible
conclusions one could draw at this point. One possibility is
that the quasiclassical expansion breaks down for $m_H$ near
 $M_W$.  The one-loop effective coupling is of order one in this
case, and one should certainly study higher loop contributions
to the effective action.

 The other logical possibility  would be a substantial cancellation
 between different terms of the derivative expansion such that
the correction to the leading quasiclassical term remains
small, but only the derivative expansion breaks down.
Since we could not carry out the heat-kernel expansion
to sufficiently high order we cannot decide this point.

For the problem of washing-out the baryon asymmetry
this may not be so relevant, because the sphaleron rate
 is far too high in the experimentally allowed mass range.
In the region where the rate is small enough for a
freeze-out of a baryon asymmetry the one-loop correction
is very  small. However, this mass range is excluded by
LEP data. Therefore, the baryon asymmetry of the universe
 cannot be understood in the minimal standard model with
one Higgs doublet just as a B+L asymmetry.  The sphaleron
 rate in the experimentally allowed Higgs mass range is so
much larger than the level required for B+L freeze-out that
higher loop, or non-perturbative effects should not be expected
 to rescue the baryon asymmetry.  Extensions of the model have
 to be considered.

The major result of this paper is a systematic treatment of
radiative corrections
to the sphaleron rate close to the critical temperature
where washing-out of a
baryon asymmetry would be most severe. Previous studies
 of
radiative corrections do not apply in this
temperature range because their effective 3-dimensional
gauge coupling diverges at
the lower critical temperature, and the whole scheme
breaks down (no matter how
small $\lambda$ is). This does not mean that sphaleron
 transitions actually become
unsuppressed at $T_0$, but simply arises from choosing
 an unfortunate scheme for
defining the coupling, not taking into account that we deal
 with a first order transition.
Qualitatively, this is quite obvious.  We have given a systematic
framework for doing
 the quasiclassical expansion in this case.
For $T_0 < T < T_c$ there are no problems  with a complex
perturbative potential.
In the  't Hooft - Feynman  background gauge all fluctuations
 including Goldstone and Higgs fields have
positive mass squared for all non-zero $\phi$
(compare eq.~(\ref{3.13})),
in this temperature range.

Similar considerations apply to the bounce
solution describing
the formation of critical bubbles at the phase
transition. In this case the $\phi^3$
term is even more  important because no
bounce solution
exists without it.  Our approach will be applied to
bubble formation  in a forthcoming publication
\cite{36}.

\acknowledgments
We would like to thank Ch. Wetterich for useful
discussions, and J. Baacke for a number of
clarifying communications.

\newpage

\begin{figure}
\caption{The accessible range of  $\lambda_T/g^2$
at $T_0$, for various values of the top
quark mass.}
\label{fig1}
\end{figure}

\begin{figure}
\caption{$A(\lambda_T/g^2,\epsilon(T))$ as function of $\lambda_T/g^2$.}
\label{fig2}
\end{figure}

\begin{figure}
\caption{Various contributions to the sphaleron action.}
%as explained in the
%Figure}
\label{fig3}
\end{figure}

\begin{figure}
\caption{$S_{FF}^{(1,\protect\beta)}$  relative to the leading contribution.}
\label{fig4}
\end{figure}

\newpage
\begin{table}
\caption{
The first few operators required for the proper
 time expansion (3.7) are evaluated
for the sphaleron background field, for two values of $\frac{\lambda_T}{g^2}$.
The operators $V_i$
are the ones listed in Appendix A of ref. [16].
% \cite{14}. Das geht nich1t automatisch !!!!!!!!!!!!
}
\begin{eqnarray*}
V_n&=&\int\frac{d^3\xi}{2\pi}(1-2\phi^+\phi)^n,\quad  n=1,2,3\nonumber\\
V_4&=&\int\frac{d^3\xi}{4\pi}(\partial_i(\phi^+\phi))^2,\nonumber\\
V_5&=&\int \frac{d^3\xi}{4\pi}F^a_{ij}F^a_{ij}\nonumber\\
V_6&=&-\frac{1}{4!}\int \frac{d^3\xi}{4\pi}
\epsilon^{abc}F^a_{ij}F^b_{jk}F^c_{ki}
\nonumber\\
V_7&=&\int\frac{d^3\xi}{4\pi}\phi^+\phi(D_k\phi)^+(D_k\phi)\nonumber\\
V_8&=&\int\frac{d^3\xi}{4\pi}\phi^+\phi F^a_{ij}F^a_{ij}\end{eqnarray*}

\begin{tabular}{cdd}
$V_i$&$\lambda/g^2=0$.$01/8$&$\lambda/g^2=1/8$\\
\tableline
1& 1253.0&14.56\\
2& 69.45& 3.152\\
3& 17.22& 1.450\\
4& 0.3956& 0.2467\\
5& 3.225& 4.036\\
6& 0.0225& 0.04536\\
7& 0.1563& 0.1800\\
8& 0.4709& 0.9387\\
\end{tabular}
\label{tab}
\end{table}

\end{document}